\documentclass[aps,prl,showpacs,twocolumn,preprintnumbers,amsmath,amsfonts,bm]{revtex4}
\usepackage[dvips]{graphicx}
\usepackage{dcolumn}

\begin{document}

\title{Long-range memory elementary 1D cellular automata:  Dynamics and nonextensivity}

\author{Thimo Rohlf$^{1}$ and 
Constantino Tsallis$^{1,2}$ 
}
\affiliation{
$^1$Santa Fe Institute, 1399 Hyde Park Road, Santa Fe, NM 87501, USA \\
$^2$Centro Brasileiro de Pesquisas Fisicas, 
Xavier Sigaud 150, 
22290-180 Rio de Janeiro-RJ, Brazil \\
}
\date{\today}

\begin{abstract}
We numerically study the dynamics of elementary 1D cellular automata (CA), where the
binary state $\sigma_i(t) \in \{0,1\}$ of a cell $i$ does not only depend on the states in its local neighborhood at time $t-1$, but also on the memory of its own
past states $\sigma_i(t-2), \sigma_i(t-3),...,\sigma_i(t-\tau),...$. We assume that the weight of this memory decays
proportionally to $\tau^{-\alpha}$, with $\alpha \ge 0$ (the limit $\alpha \to \infty$ corresponds to the usual CA).
Since the memory function is summable for $\alpha>1$ and nonsummable for $0\le \alpha \le 1$, we expect pronounced 
changes of the dynamical behavior near $\alpha=1$. This is precisely what our simulations exhibit, particularly for the time evolution of the Hamming distance $H$ of initially close trajectories. 
We typically expect the asymptotic behavior $H(t) \propto t^{1/(1-q)}$, where $q$ is the entropic index associated with nonextensive statistical mechanics. In all cases, the function $q(\alpha)$ exhibits a sensible change at $\alpha \simeq 1$.
 We focuse on the 
 class II rules 61, 99 and 111. For rule 61, $q = 0$ for $0 \le \alpha \le \alpha_c \simeq 1.3$, and $q<0$ for $\alpha> \alpha_c$, whereas the opposite behavior is found for rule 111. For rule 99, the effect of the long-range memory on the spread of damage is quite dramatic. These facts point at a rich dynamics intimately linked to the interplay of local lookup rules and the range of the memory. Finite size scaling studies varying system size $N$ indicate that the range of the power-law regime for $H(t)$ typically diverges $\propto N^z$ with $0 \le z \le 1$. Similar studies have been carried out for other rules, e.g., the famous ``universal computer" rule 110.
\end{abstract}


\maketitle

Cellular Automata (CAs) have a long tradition as models of complex, emergent space time dynamics encoded in simple, local rules. Originally introduced
by \emph{von Neumann} as a theroretical framework for self-replication \cite{vonNeumann52}, it was realized that they could serve as models for much broader
classes of phenomena, including even universal computation (e.g. in Conway's "game of life" \cite{Berlekamp82}). In the context of statistical mechanics, detailed
studies on the simplest class of CA, elementary 1D CA with $k=2$ states and 3-cell neighborhood, where carried out in the 1980's.
Wolfram \cite{Wolfram83,Wolfram84} developed a qualitative classification scheme of the $2^{2^3} = 256$ elementary CA rules (only 88 fundamentally inequivalent) that distinguished four different 'complexity classes' of their dynamics
(class I: fixed-point attractors, class II: space-time periodic attractors (limit cyles), class III: aperiodic space-time chaos, class IV: 'complex'
dynamics (travelling, localized aperiodic structures on regular background)). It was shown that rule 110, a member of class IV, is a universal computer
in the Turing sense \cite{WolframCook2002_2004}. Many attempts where undertaken to obtain a more quantitative characterization of CA dynamics, e.g. mean field models
\cite{Schulman78,Wolfram83}, local structure theory \cite{Gutowitz87}, quantification of pre-images \cite{GomezSoto2006} and relating certain class III rules 
to low-dimensional deterministic chaos \cite{Andrecut2001}. The relation between the different classes,
as well as the connection of these discrete, dynamical systems to other fields in statistical mechanics, however, still remained obscure, leaving open
the question of whether there might exist a more general theoretic framework to describe CA dynamics. Fractal patterns and strong spatial and temporal
correlations observed in CA dynamics for certain rules, as well as the absence of a Hamiltonian, seem to suggest a natural connection to \emph{nonextensive statistical mechanics} and the related 
concept of $q$-entropies (with $q \ne 1$) \cite{Tsallis88,GellMannTsallis04,BoonTsallis05}, rather than the traditional Boltzmann-Gibbs (BG) framework (i.e., the limiting case $q =1$). 
Whereas systems in the BG class are characterized by strong sensitivity on initial conditions with trajectories typically diverging {\it exponentially} with time (hence strong mixing in phase space, and ergodicity), systems with $q \ne 1$ typically show a weaker dependence on initial conditions, e.g.
trajectories diverging asymptotically as a {\it power} of time. Therefore, BG systems may be characterized as 'memoryless systems', whereas the systems studied in
nonextensive statistical mechanics often exhibit long-range memory.

\begin{figure}[htb]
\begin{center}
\resizebox{55mm}{!}{\includegraphics{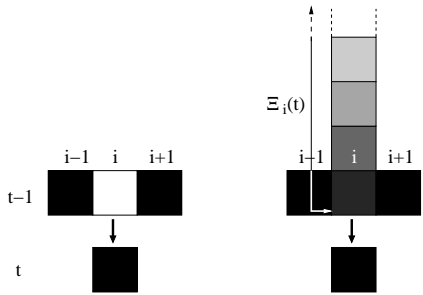}}
\end{center}
\caption{\small {\em Left panel}: in conventional CA, the state of cell $i$ at time $t$ depends only on the states in its local neighborhood at time $t-1$. {\em Right panel}: in $\alpha$-CA, the state of cell $i$ at time $t$ depends on the states of its neighbors at time $t-1$ and the (infinite) memory $\Xi_i(t)$ of its own past states (grey shading indicates power-law decay of the weighting of memory states).}
\label{updatescheme}
\end{figure}

In this paper, we first define a new model by explicitely introducing long-range memory. In our model, the binary state $\sigma_i(t) \in \{0,1\}$ of a cell $i$ does not only depend on the states in its local neighborhood at time $t-1$, but also on the memory of its own past states $\sigma_i(t-2), \sigma_i(t-3),...,\sigma_i(t-\tau),...$. We assume that the weight of this memory decays
proportionally to $\tau^{-\alpha}$, with $\alpha \ge 0$. The case  $\alpha \to \infty$ corresponds to the usual CA, where the states at time $t$ are determined solely by the neighbor states at time $t-1$. This case is compared to the generalized update scheme in Fig. \ref{updatescheme}.
We then study its sensitivity to  initial conditions. More precisely, we determine the $\alpha$-dependence of the entropic index $q$ (sometimes noted $q_{sen}$ in the literature, where {\it sen} stands for {\it sensitivity}). By analogy with what occurs in simple dissipative maps at the edge of chaos \cite{Robledo},  in vanishing Lyapunov exponent conservative maps  \cite{Casati}, and in self-organized critical models \cite{SOC}, one expects the sensitivity to the initial conditions (the Hamming distance in the present case) to be asymptotically proportional to $t^{1/(1-q)}$ with $q<1$. 
While for some elementary CA, e.g. rule 110, long range memory seems to emerge as a result of purely local
update rules, the way this happens (and, why it does \emph{not} happen for most other rules) is poorely understood. Our approach, that contains the conventional CA
scheme as a limit case, aims at providing a generalized framework to address such questions, and also relating them to nonextensive statistical mechanical concepts.

\emph{Model}. 
Consider a 1D cellular automata, consisting of $N$ cells in a line, with periodic boundary conditions. The state space of the system is defined by
 $\vec{\sigma} \in\{1,0\}^N$.
Each cell $i$ is updated in parallel according to the following update rule:
\begin{equation}
\sigma_i(t) = f[ \sigma_{i-1}(t-1), \Theta(\Xi_i(t) - 1/2 ), \sigma_{i+1}(t-1) ] ,
\end{equation}
where $\Xi_i(t)$ is defined by:
\begin{equation}
\Xi_i(t) = \lim_{T \to \infty}\sum_{\tau = 1}^{T}\frac{\sigma_i(t-\tau)}{\tau^{\alpha}}  /  \sum_{\tau = 1}^{T}\frac{1}{\tau^{\alpha}} 
\end{equation}
with $\alpha \in [0,\infty)$. $\Theta(x)$ is the Heaviside step function, i.e., due to the normalization of $\Xi \in [0,1]$, it returns either $0$
or $1$, depending on $\Xi_i(t) - 1/2$ being smaller or larger than zero, respectively. 
$f$ is one of the well-known 256 elementary local update rules for neigborhood 3, binary CA. However, $\Xi_i(t)$ introduces long-range memory, 
decaying with a power $\alpha$ of (discrete) system time. Notice that for $\alpha \to \infty$, this maps on the conventional 1D CA without memory.
In practice (i.e. simulations), $T$ does not go to infinity, but rather is fixed to some large finite value ($T$ up to 960).
Let us briefly discuss the spirit of this approach. While it would be perfectly reasonable to choose any other functional form for memory decay (e.g., an
exponential), for our purposes, the one chosen in Eq. (2) appears to be the most sensible: it becomes, for arbitrary configurations and $T \to \infty$, nonsummable for $0 \le \alpha \le 1$, hence, we
expect a transition from a short-range memory phase to a long-range memory phase if we approach the critical value $\alpha_c$ from above. Consequently, it is the simplest
functional form for which we may expect a qualitative change in dynamics for a finite value of the control parameter $\alpha$, while keeping the update rule
$f$ constant.

\begin{figure}[htb]
\begin{center}
\resizebox{80mm}{!}{\includegraphics{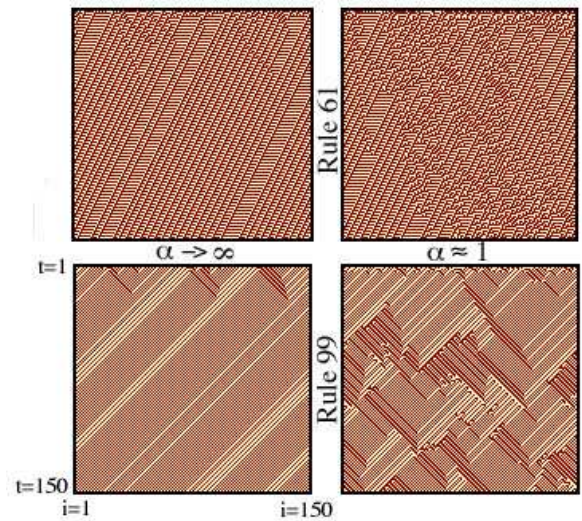}}
\end{center}
\caption{\small
Space-time plots starting from random
 initial configurations of conventional CA, i.e. $\alpha \to \infty$ (left panels), and $\alpha = 1.2$ for rule 61, $\alpha =1$  for rule 99 (right panels).
  States $\sigma_i = 0$ are shown yellow, $\sigma_i =1$ in red.}
\label{rul61_99dyn}
\end{figure}

\begin{figure}[htb]
\begin{center}
\resizebox{65mm}{!}{\includegraphics{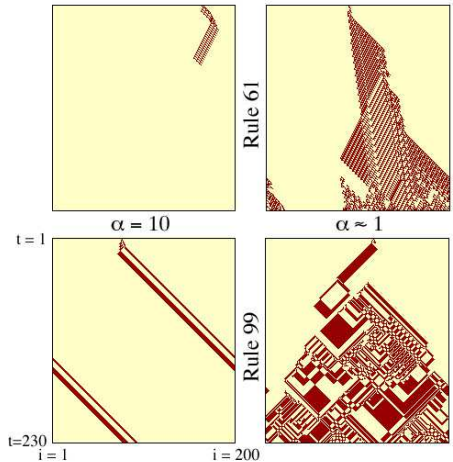}}
\end{center}
\caption{\small Difference patterns for CA with initial configurations differing in one randomly chosen bit. Cells with different states in both configurations at time
$t$ are shown in red. Right panels:  $\alpha = 1.2$ ($\alpha = 1$) for rule 61 (rule 99). 
 }
\label{rul61_99spread}
\end{figure}

\emph{Results}. We have carried out extensive numerical simulations for many of the 256 elementary CA rules. It turns out that class I and class III CA
 ususally are very insensitive to long range memory, i.e. trajectories of $\alpha$-CAs still
converge to fixed points (class I) or cover a huge proportion of the state space in a 'chaotic' manner (class III), independent of  $\alpha$.  However, quite pronounced changes in system dynamics are found for several class II and class IV rules.
Here, we focuse our discussion on three rules that where classified as class II, namely rule 61, 111 and 99. In the conventional updating scheme, 
all three rules behave very similarly, i.e. their dynamics converges to space-time periodic, checker-board like patterns traveling over the lattice (compare
left panels of Fig.2). Near $\alpha_c \approx 1$, however, profound changes in the dynamics are observed: due to increased sensitivity
towards initial conditions, we find complex, fractal spatio-temporal patterns (perturbations) travelling on a regular background. Whereas for small $N$
these perturbations tend to die out after a finite number of updates, for large $N$ they appear to be (asymptotically) stable.
Similar striking changes are found in the time evolution of \emph{difference patterns}, when compared to the coventional update scheme (Fig. \ref{rul61_99spread}).

\begin{figure}[htb]
\begin{center}
\resizebox{85mm}{!}{\includegraphics{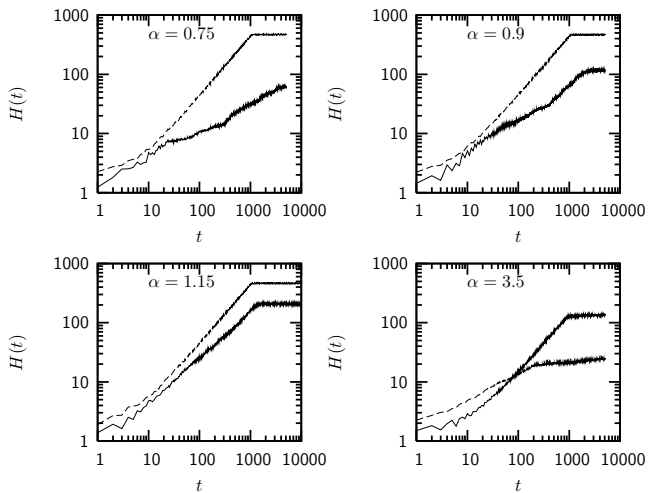}}
\end{center}
\caption{\small Time dependence of the Hamming distance $H(t)$ for rule 111 (filled curves) and rule 61 (dashed curves) for typical values of $\alpha$ and $N = 1000$, ensemble averages over 200 different initial conditions. All curves approximately follow straight lines in the log-log plots for two decades or more, indicating power law behavior. Memory size $T=320$.}
\label{hammcompare}
\end{figure}

\begin{figure}[htb]
\begin{center}
\resizebox{85mm}{!}{\includegraphics{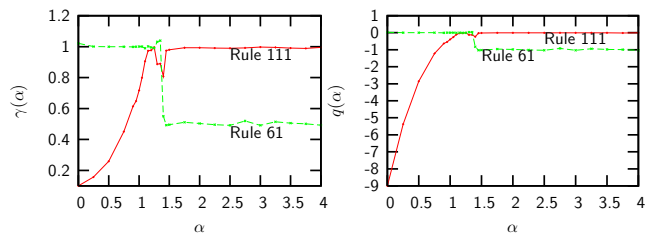}}
\end{center}
\caption{\small The $\alpha$-dependence of $\gamma$ (left) and $q$ (right)  for rules 61 and 111. Memory size $T=320$.}
\label{alphscale}
\end{figure}

Let us now systematically study sensitivity towards initial conditions. 
We measured the time evolution of the Hamming distance 
$H(t)$ of initially close trajectories for different CA rules and different values of the memory parameter $\alpha$. 
Typically, we find asymptotic scaling proportional to
a power-law with an exponent $\gamma$ that is a function of $\alpha$:
\begin{equation}
H(t) \propto \,t^{\gamma(\alpha)} \quad\mbox{for}\,\, t > t_0,
\end{equation}
where $t_0$ is small, e.g. in the order of $10$ for rule $61$ if the initial configurations differ in one (randomly chosen) bit.
Fig. \ref{hammcompare} compares $H(t)$ of rules $61$ and $111$, for different values of $\alpha$. In all four cases shown,
we find power-law scaling for two decades or more, however, the functional dependence of the slope $\gamma(\alpha)$ is obviously
very different for both rules. This is also evident in Fig. \ref{alphscale}: whereas for rule 61, $\gamma$ is very close to unity
for all $\alpha < \alpha_c^{61} \approx 1.4$ and shows a steep, step-like transition to $\gamma \approx 1/2$ for $\alpha > \alpha_c^{61}$,
rule 111 shows a (nearly) opposite bebavior - for $\alpha > \alpha_c^{111}$, one has $\gamma \approx 1$, for $\alpha$ below the critical point,
a smooth transition to $\gamma(0) \approx 0$ is found. The exact value of $\lim_{\alpha \to 0}\gamma(\alpha)$ is, however, still ambiguous, as the measured
slope in this limit is affected by finite size effects of the memory, tending to overestimate $\gamma$. In simulations in this regime we used memory sizes up to $T = 960$; the smooth decay of $\gamma(\alpha)$
strongly suggests that $\lim_{\alpha \to 0} \lim_{T \to \infty}\gamma(\alpha,T) = 0$. 

In nonextensive statistical mechanics, $q$ is a measure for the sensitivity of a dynamical system towards variation of initial
conditions; it is closely related to the divergence of trajectories in the phase space of the system. In case of divergence
of trajectories described by a power-law $\propto t^{\gamma}$, we have the relation
\begin{equation} \gamma = 1/(1-q) \Leftrightarrow q = 1 - 1/\gamma.\end{equation}
The functional behavior of $q(\alpha)$ for rules 61 and 111 is plotted in Fig. \ref{alphscale} (left panel). For both rules, we find
strong deviations from $q=1$ (i.e. the exponential divergence typical of classical BG statistics): in the case of rule 61, one finds $q \approx 0$ for $\alpha < \alpha_c^{61}$ and $q \approx -1$
for larger values of $\alpha$; for rule 111, $q$ is strongly negative below $\alpha_c^{111}$ (with a probable divergence $q(\alpha) \to -\infty$ for $\alpha \to 0$), and zero
for larger $\alpha$.

\begin{figure}[htb]
\begin{center}
\resizebox{85mm}{!}{\includegraphics{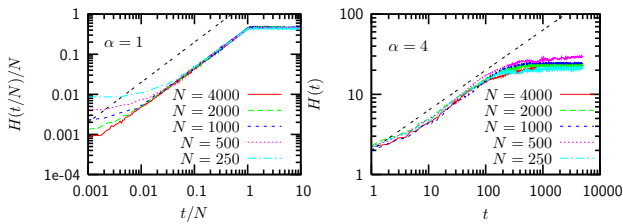}}
\end{center}
\caption{\small Time dependence of the Hamming distance for rule 61 at different sizes $N$. {\it Left:} $t$ and $H$ are rescaled by $N$. The collapse of the curves indicates that the range of the power-law regime for $H(t)$ diverges with system size $N$ (straight dashed line with $\gamma = 1$ shown for eye guidance). {\it Right:} For $\alpha \ge 1.4$ curves collapse without
rescaling (straight dashed line has slope $\gamma = 1/2$). 
Memory size $T=320$.}
\label{collapsed}
\end{figure}

Last, let us look at the critical exponents $z(\alpha)$ that describe the divergence of the range of the power-law regime with system size $\propto N^z$. 
For rule 111, one finds
$z = 1$ independent from $\alpha$, whereas for rule 61, one finds $z(\alpha) \approx 1$ for $\alpha < \alpha_c^{61}$ and $z(\alpha) \approx 0$ above $\alpha_c^{61}$
(Fig. \ref{collapsed}). This seems to suggest that the transition at  $\alpha_c$ is second-order like for rule 111 and first-order like for rule 61, which is also confirmed by the step-like discontinuity in $\gamma(\alpha)$ in the latter case and the smooth descend of this quantity below $\alpha_c$ in the former case (Fig. \ref{alphscale}, right panel).

\emph{Discussion}. We demonstrated that long-range memory with weights decaying $\propto \tau^{-\alpha}$ leads to completely unexpected, exciting new dynamical phenomena
in elementary 1D cellular automata. In particular, we showed for the three class II rules 61, 99 and 111 that the sensitivity towards initial conditions,
measured in terms of the divergence of the Hamming distance between initially close configurations $H(t) \propto t^{\gamma}$, and the associated entropic index
$q$, shows pronounced transitions at critical values $\alpha_c$ slightly above one. Interestingly, the behavior of $q(\alpha)$ is strikingly different for rules
61 and 111: whereas the former shows a step-like transition from $q = 0$ to $q = -1$ at  $\alpha_c$ with increasing $\alpha$, the latter has $q=0$ above $\alpha_c$ and
a gradual divergence to $-\infty$ when alpha approaches zero. Let us first comment the fact that the transition takes place at $\alpha_c > 1$, although the memory function
in Eq. (2) becomes nonsummable for $\alpha \le 1$; this is probably due to two effects: first, the effective dimension of the system is slightly \emph{larger} than one,
because not only the 1D memory strings, but also the local neighborhoods of cells are considered; second, there is a non-trivial interplay between the memory and
the update rules, leading e.g. to considerable fluctuations of $q(\alpha)$ at $\alpha_c$ that appear to be conserved even in the limits $N \to \infty$ and $T \to \infty$ (compare Fig. \ref{alphscale}). However, a more detailed analysis is necessary to quantify the influence of both effects.
Considering the fact that, without long-range memory,
 both rules converge to very similar dynamical attractors (and hence were both assigned
to class II by Wolfram), these striking differences are really a surprise. Remarkably, the different behavior wrt $q$ is still conserved in the limit of large $\alpha$,
where our new, generalized scheme maps on the conventional CA without memory. This may indicate that the application of these concepts derived from nonextensive statistical mechanics could help to refine existing CA classification schemes.
In particular, complex dynamical behavior, as e.g. found for the famous 'universal computer rule' 110, depends on a delicate interplay between memory and information spreading, however, this is poorly understood for most CA. Our generalized scheme may help to address this interesting \emph{reverse problem}, i.e. the question of how memory emerges for certain CA rules, and why it does not for others (we will report details elsewhere).

We acknowledge interesting discussions with J.M. Gomez Soto, as well as financial support by SI International and AFRL (USA).

\end{document}